\documentclass[aps,prl,twocolumn,groupedaddress]{revtex4}

\usepackage{color}
\usepackage{graphicx}
\usepackage{amsmath, amsthm,amssymb}
\begin{document}
%\preprint{}

\title{ A Universal Magnetic Helicity Integral}


\author{Gunnar Hornig} 
\affiliation{Division of Mathematics, \\ University of Dundee 
  Dundee DD1 4HN, \\ United Kingdom}%
\email{gunnar@maths.dundee.ac.uk}
\homepage{http://www.maths.dundee.ac.uk/mhd/}
%\thanks{}
%\affiliation{}


\date{\today}

\begin{abstract}
A  magnetic helicity integral is proposed  which can be applied to domains which 
are not magnetically closed,  i.e.~have a non-vanishing normal component of the 
magnetic field on the boundary. In contrast to the relative helicity integral, 
which was previously suggested for magnetically open domains, it does not rely on 
a reference field and thus avoids all problems related to the choice of a 
particular reference field. Instead it uses a gauge condition on the vector 
potential, which corresponds to a particular topologically unique closure of the 
magnetic field in the external space. The integral has additional elegant 
properties and is easy to compute numerically in practice.  For magnetically 
closed domains it reduces to the classical helicity integral.

\end{abstract}

\pacs{52.30.Cv, 95.30.Qd,  03.50.De, 02.40.M} 
        %  Classical electromagnetism 03.50.De
        % Magnetohydrodynamics in Plasmas 52.30.Cv
        %  Global differential geometry  02.40.M, 
        %  solar magnetic fields
        %  96.60.H, Magnetohydrodynamics in astrophysics, 95.30.Qd

%\keywords{}

\maketitle

\section{Introduction}
Magnetic helicity  is an important quantity in describing the structure and 
evolution of magnetic fields in many fields of physics, in particular in plasma 
physics and astrophysics. It was introduced to Plasma Physics  by H.K.~Moffatt in 
\cite{Moffatt69} and was originally defined as an integral over a magnetically closed volume, i.e.~a 
volume for which  the normal component of the magnetic field (${\bf B}$) vanishes on 
the boundary $\partial V$:
\begin{equation}
H({\bf B}) := \int_{V} {\bf A} \!\cdot\! {\bf B} \ dV \ ; \quad B_{n}= {\bf B}\!\cdot\! 
{\bf n}\vert_{\partial V}=0 . \label{totalhelicity}
\end{equation}
Here ${\bf A}$ is the vector potential, $\nabla \times {\bf A} = {\bf B}$, 
of the magnetic field.
The integral measures - roughly speaking -  the Gaussian linkage of magnetic flux 
within $V$. More precisely, it is the asymptotic linking number of pairs of field 
lines averaged over the volume  \cite{Arnold86}. It is an  important property  of 
this integral that we can derive an equation of continuity for the helicity 
density, which uses only the homogenous Maxwell's equations (here  
${\bf E}= - \partial_t {\bf A}- \nabla \phi$ is the electric field):
\begin{equation}
\partial_{t} ({\bf A} \!\cdot\! {\bf B}) + \nabla \!\cdot\! \left( \phi {\bf B} 
+ {\bf E} \times {\bf A} \right) = -2 \ {\bf E} \!\cdot\! {\bf B}  \  . 
\label{helevolu}
\end{equation}
It can be shown that the integral is a topological invariant, i.e.~it does not change 
under a deformation of the field within $V$, as given for instance by the motion 
of a magnetic field embedded  in an ideal plasma, satisfying (${\bf v}$ is the plasma 
velocity)
\begin{equation}
{\bf E} + \mathbf{v}\times\mathbf{B}  =  0   \label{idealevolu} \ .
\end{equation}
Under such a condition,  (\ref{helevolu}) becomes
\begin{equation}
\partial_{t} ({\bf A} \!\cdot\! {\bf B}) + \nabla \!\cdot\! \left( (\phi -{\bf 
v}\!\cdot\! {\bf A}) {\bf B} + {\bf v}  {\bf A}\!\cdot\!{\bf B} \right)  =   0   \ 
, \label{idealevoluhel} 
\end{equation}
so that integrating over a volume with ${\bf v}\!\cdot\! {\bf n}=0$ on the 
boundary (or more generally a comoving volume) results in
\begin{eqnarray}
\frac{d}{dt} \int_{V} {\bf A} \!\cdot\! {\bf B} \ dV & = & \int_{V} 
\partial_{t} ({\bf A} \!\cdot\! {\bf B}) + \nabla  \!\cdot\!  ({\bf v} \  {\bf 
A}\!\cdot\!{\bf B} ) \ dV  \nonumber \\
& =&  - \int_{\partial V}\!\!\! (\phi -{\bf v}\!\cdot\! {\bf A}) {\bf 
B}\!\cdot\!{\bf n}  \, da  =  0. \label{changehel2}
\end{eqnarray}
Moreover, the total helicity is often an approximate invariant for non-ideal 
plasmas \cite{Berger84}, and is therefore a valuable tool in determining the evolution 
of many 
technical and natural plasmas. One of the earliest results was the prediction of 
the relaxed state of a Reversed-Field Pinch \cite{Taylor:RelaxationReconnection}, 
but there are many more applications (see  \cite{HelicitySpaceLab} for an 
overview).

However, the boundary condition $B_{n}=0$ on the 
integral (\ref{totalhelicity}), which is necessary to ensure gauge invariance, 
means that it can not  be applied to cases where the magnetic field crosses the 
boundary. Typical examples are the vacuum vessels of laboratory plasmas where an 
external magnetic field crosses the boundaries, or the atmospheres of stars or 
planets, where the studied volume is usually bounded by the surface of the body, 
through which the magnetic field emerges. 

In such cases it was previously necessary to resort to the calculation of the 
relative helicity, i.e.~the helicity was calculated with respect to a reference 
field ${\bf B}_{\rm ref}$ satisfying the same boundary conditions. One can prove  
\cite{FinnAntonsen,Berger1984} that for an arbitrary closure of the magnetic field 
outside $V$, denoted by ${\bf B}_{\rm ext}$, the relative helicity 
\begin{eqnarray} 
H({\bf B} \vert {\bf B}_{\rm ref}) & = & H({\bf B}+{\bf B}_{\rm ext})-H({\bf 
B}_{\rm ref}+{\bf B}_{\rm ext}) \\
              & = & \int ({\bf A} + {\bf A}_{\rm ref}) \cdot ({\bf B} -{\bf 
B}_{\rm ref}) \ dV \ ,
\end{eqnarray}
is actually independent of the external closure of the field.  The reference field 
is in most cases choosen to be  a potential field (see 
e.g.~\cite{BergerRuzmaikin:HelicityProduction}) since a potential field is easy to 
compute and  physically distinguished as the lowest energy state compatible with 
the boundary conditions. The introduction of a reference field, however, not only 
complicates the calculation of magnetic helicity, but also complicates  its 
already difficult  interpretation.  For instance, the question arises as to 
whether a change of relative helicity in a volume has a physical meaning, or 
whether it is only due to our particular choice of reference field.

In this contribution it is proposed to replace  the reference field by a more 
general boundary condition on the vector potential and it is shown that this leads 
to a well defined quantity.

\section{Definition}
The  volume  $V \subset I\!\!R^{3}$ we refer to is assumed to be simply connected 
and without cavities, i.e.~it has vanishing first and second Betti numbers. The 
first condition ensures that the vector potential is unique up to a gradient of a 
function, while the second implies that a vector potential always exist 
(alternatively we can require $\int {\bf B}\!\cdot\!{\bf n} da=0$ over the 
boundary of any cavity). More general volumes such as a solid torus  can be 
considered as well, if additional constraints are included. For a solid torus, for 
example, we have to impose $\int {\bf A}\!\cdot\! d{\bf l} =0$  around the hole of 
the torus.

\paragraph{Definition:} {\it 
The  universal magnetic helicity of a magnetic field in a simply connected volume 
$V \subset I\!\!R^{3}$  is defined as
\begin{equation}
H_{V}({\bf B}) = \int_{V} {\bf A} \!\cdot\! {\bf B} \ dV \  \mbox{with} \  
\nabla_T \!\cdot\! {\bf A}_T =0 \ \mbox{on} \ \partial V. 
\label{generalisedhelicity}
\end{equation}
}

Here  $\nabla_T \!\cdot\! {\bf A}_T =0$ is the divergence  of the tangential 
component of ${\bf A}$ on  the boundary $\partial V$, i.e.~the divergence is taken 
with respect to the boundary coordinates only. For an explicite calculation  we can
choose an orthogonal curvilinear coordinate system $(u_1,u_2,u_3)$, defined locally 
on the boundary such that the unit vector ${\bf u}_3$ coincides with ${\bf n}$ . 
Then $\{{\bf u}_1, {\bf u}_2 \}$ span the tangent space of 
the boundary and ${\bf A}_T$ is represented as ${\bf A}_T = A^1 {\bf u}_1 + A^2 {\bf 
u}_2$
Denoting the scaling factors $h_i = \| \frac{\partial {\bf r}}{\partial u_i} \|$ the 
divergence reads
\begin{equation} \nabla_T \cdot {\bf A}_T = \frac{1}{h_1 h_2} 
\left(\frac{\partial}{\partial u_1}h_2 A^1 
+ \frac{\partial}{\partial u_2}h_1 A^2 \right) \label{expdiva}  \ . \end{equation}

In order for the quantity $H_{V}({\bf B})$ to be well defined, we have to prove 
firstly that it is not gauge dependent, and secondly  that the boundary condition 
can be satisfied for any field. 
\paragraph{Gauge invariance:} First note that the curl of ${\bf A}_{T}$ on the boundary, that is in a two-dimensional surface is a scalar.  Using the coordinate  system from above it reads
\begin{equation}
\nabla_T \times {\bf A}_T = \frac{1}{h_1 h_2} 
\left(\frac{\partial}{\partial u_1}h_2 A^2 
- \frac{\partial}{\partial u_2}h_1 A^1 \right) = B_{n} \ .  \label{2drot}
\end{equation}
Together with the boundary condition  $ \nabla_T \!\cdot\! {\bf A}_T =0$, it  uniquely determines ${\bf A}_T$, since a gauge ${\bf A} \rightarrow 
{\bf A} + \nabla \lambda$ consistent with the boundary condition requires 
$\nabla_T \cdot (\nabla \lambda)_T$ $=$ $\triangle_T \lambda = 0$ on the boundary, which is a  closed 
manifold, and therefore $\lambda\vert_{\partial V}=constant$.  This means 
that $\lambda$ is constant on the boundary, but it will in general vary inside 
$V$. 
However, a gauge transformation  ${\bf A} \rightarrow 
{\bf A} + \nabla \lambda$ with $\lambda\vert_{\partial V}=const.$ conserves the helicity 
integral:
\begin{eqnarray} 
H_{V}({\bf B}) & \rightarrow & H_{V}({\bf B}) + \int_{V} \nabla \lambda \cdot {\bf B} 
\!\cdot\! {\bf n}\, da \nonumber\\
\int_{V} \nabla \lambda \cdot {\bf B} &=& \int_{\partial V} \lambda {\bf B} \!\cdot\! {\bf 
n}\, da 
=  \lambda \int_{\partial V} {\bf B} \!\cdot\! {\bf n} \, da = 0  \ .
\end{eqnarray}

\paragraph{Existence:}
Starting with an arbitrary vector potential for ${\bf B}$ in $V$,  with $ 
\nabla_T \!\cdot\! {\bf A}_T  =\rho$, we note that it is possible to find a 
function $\lambda(u_{1},u_{2})$ on the boundary with 
$\triangle_T \lambda  = \rho$.  Here $u_{1}$ and $u_{2}$ are boundary coordinates as  defined above.  This function  can be extended to a 
function on all of $V$, for instance,  by letting it smoothly fall off to zero 
within an $\epsilon$-neighbourhood of the boundary $\lambda({\bf x}) :=  \lambda(u_{1},u_{2}) f(u_{3}) $ with 
\begin{equation}
f(u_{3}) := \left\{    \begin{array}{ll}
      \exp(-u_{3}^{2}/(\epsilon^{2}-u_{3}^{2})) & u_{3} \le \epsilon \\
      0 & \mbox{elsewhere} \\
   \end{array}  \right. \label{fdef}
\end{equation}
 Thus ${\bf \hat A}= {\bf A} - \nabla \lambda$ has the 
desired property $ \nabla_T \!\cdot\! {\bf \hat A}_T  =0$.  Note that this proof also shows how to satisfy the boundary condition in practice. In particular there exist standard numerical routines to solve  $\triangle_T \lambda  = \rho$ on arbitrary boundaries.

\section{Properties}
Showing that the universal helicity is well defined is not enough to justify its 
name. It must also reduce to the total helicity (\ref{totalhelicity}) for the case 
of a vanishing normal component of ${\bf B}$ on the boundary. Since 
(\ref{generalisedhelicity}) includes the case of vanishing $B_{n}$ this is obviously the case. 

In addition, we can prove that it is a topological invariant for any deformation 
of the magnetic field inside $V$, i.e.~for any deformation, which leaves the 
boundary unaffected: ${\bf v}=0\vert_{\partial V}$.  Evaluating (\ref{idealevolu}) 
on the boundary gives ${\bf E}_T= -\partial_{t }{\bf A}_T - (\nabla 
\phi)_T =0$. Since ${\bf A}_T$ is determined solely by ${\bf B}_{n}$, which  
is constant in time, we obtain $\partial_{t }{\bf A}_T=0$ and hence $\phi 
=const.$ on the boundary. Thus (\ref{changehel2}) vanishes, now due to $\phi 
=const.$ and ${\bf v}=0\vert_{\partial V}$ instead of due to $B_{n}=0$.

\begin{figure}[h]
\includegraphics[width=6cm]{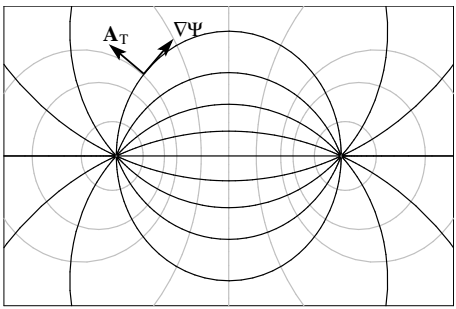}
\caption{An example of two adjacent regions of positive and negative 
${\bf B} \!\cdot\! {\bf n} = - \triangle_T \Psi$ \label{boundary1} on the boundary.
The grey lines are contours of the potential $\Psi$ defined by 
(\ref{orthogonalA}). They coincide with field lines of 
${\bf A}_T$. In black field lines of $\nabla \Psi$  are  shown.}
 \end{figure}

Another important property of the universal helicity integral is its additivity  
with respect to magnetic fields. The rule is the same as for the total helicity:
\begin{equation}
H_{V}({\bf B}^{a}+ {\bf B}^{b}) = H_{V}({\bf B}^{a})+ H_{V}({\bf B}^{b}) + 2 
H_{V}({\bf B}^{a}, {\bf B}^{b}),
\end{equation}
where 
\begin{equation}
H_{V}({\bf B}^{a}, {\bf B}^{b}) =\int_{V} {\bf A}^{a} \!\cdot\! {\bf B}^{b} \ dV = 
\int_{V} {\bf A}^{b} \!\cdot\! {\bf B}^{a} \ dV   \label{crossints}
\end{equation}
is the mutual or cross helicity integral. The equivalence of the two integrals in 
(\ref{crossints}) can be shown by using the condition $\nabla_T \!\cdot\! {\bf 
A}_T=0$, which implies a representation 
\begin{equation}
{\bf A}_T= \nabla \Psi \times {\bf n}.  \label{orthogonalA}
\end{equation}
In order to proof this note that  ${\bf A}_T$  has a representation as ${\bf A}_T=  {\bf a}_T \times {\bf n}$ with another vector field ${\bf a}_T$ tangential to the boundary. Comparing  (\ref{expdiva}) with (\ref{2drot}) we see:
\begin{equation} 0= \nabla_T \cdot {\bf A}_T  = \nabla_T \times {\bf a}_T  \  .
\end{equation}
Thus the field ${\bf a}_T$ is a gradient. Analogous we find
\begin{equation} B_{n} =\nabla_T \times {\bf A}_T  
= -\nabla_T \cdot {\bf a}_T = -\triangle_T \Psi 
\end{equation}
An example of ${\bf A}_T$ and the potential $\Psi$ on the boundary is shown 
in Fig.~\ref{boundary1}.

The representation (\ref{orthogonalA}) can now be used to prove that the difference 
between the two integrals in (\ref{crossints})vanishes:
\begin{eqnarray}
\int_{V} \nabla \!\cdot\! \left( {\bf A}^{a} \times {\bf A}^{b}\right) dV & = & 
\int_{\partial V}\left( {\bf A}^{a} \times {\bf A}^{b}\right) \!\cdot\! {\bf n} \ 
da  \\
 & = & \int_{\partial V}  \left( \nabla \Psi^{a} \times \nabla \Psi^{b}\right) 
\!\cdot\! {\bf n} \ da  \nonumber \\
 & = & \int_{\partial V}\left( \nabla  \times  \Psi^{a} \nabla \Psi^{b}\right) 
\!\cdot\! {\bf n} \ da  = 0.  \nonumber 
\end{eqnarray}
Here the last integral vanishes due to the application of Stokes theorem to a 
surface without boundary.

Furthermore, the universal helicity is  additive  with respect to  complementary 
volumes. Complementary  here describes  that the volumes $V^{a}$ and  $V^{b}$ are 
adjacent to one another, and that they satisfy $B^{a}_{n} = 
B^{b}_{n}$ on their common boundary, and have $B^{a}_{n} = B^{b}_{n}=0$ on all  other 
boundaries.
Note that this still assumes that the volumes are simply connected and have no 
holes. Thus, the total volume $V^{a} \cup V^{b}$ has vanishing normal magnetic 
field on its boundary, and we can calculate its (classical)  total helicity 
\begin{equation}
H_{V^{a} \cup V^{b}}({\bf B}^{a}+ {\bf B}^{b}) =  H_{V^{a} }({\bf B}^{a}) + 
H_{V^{b} }({\bf B}^{b}).   \label{sumvolumes}
\end{equation}
The proof relies on the fact that the total helicity on the left hand side is 
gauge invariant, so that we can choose a gauge for the vector potential such that 
$\nabla_T \!\cdot\! {\bf A}_T=0$ holds both on its boundary and on  the 
interface between $V^{a}$ and $V^{b}$.  Then the total vector potential can be 
split in two parts with ${\bf A}^{a} = {\bf A}\vert_{V^{a}}$ (${\bf A}^{a}=0$ on 
$V^{b}$) and ${\bf A}^{b}$ analogously defined. This implies (\ref{sumvolumes}).

\section{Interpretation and Example\label{interpretation}}
\begin{figure}
 \includegraphics[angle =-90,width=\columnwidth]{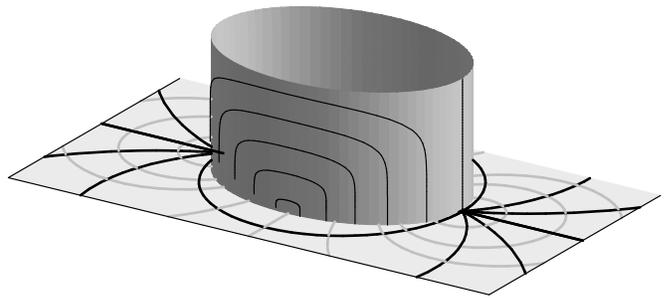}
\caption{\label{boundary2} A flux surfaces of the field ${\bf B}^b$ derived from 
(\ref{Ab}) for the same boundary condition as used in Fig.~\ref{boundary1}.}
 \end{figure}

In the following we prove that the boundary condition $\nabla_T \cdot {\bf A}_T=0$ 
 corresponds to the existence of a unique field  ${\bf B}^b$ in the exterior 
 domain $V^b$ such that ${\bf A}^b\cdot {\bf B}^b \equiv 0$. Thus the univeral 
 helicity integral $H_{V^a}({\bf B}^a)$ equals the total helicity integral of ${\bf B}^{a}$ completed by ${\bf B}^{b}$ as 
 Eq.~(\ref{sumvolumes}) shows.  In order to see this we 
 use a coordinate system as introduced for the representation (\ref{expdiva}) and 
 define ${\bf A}^b$ in a layer of thickness $\epsilon$ from the boundary with $f(u_{3})$ as defined in (\ref{fdef}):
 \begin{eqnarray}
 {\bf A}^b & = & {\bf A}_T(u_1,u_2) f(u_3)  \label{Ab} \\
 {\bf B}^b & = & B_n(u_1,u_2) f(u_3) {\bf u}^3 + \frac{\partial f(u_3)}{\partial u_3} 
\nabla \Psi(u_1,u_2)
 \end{eqnarray} 
One easily checks that this satisfies ${\bf A}^b\cdot {\bf B}^b \equiv 0$.
Note that the field ${\bf B}^b$ is tangent to surfaces spanned by ${\bf u}_3$ and 
$\nabla \Psi$. Such a surface is shown in Fig.~\ref{boundary2}. 
In the limit of $\epsilon \rightarrow 0$ the field ${\bf B}^b$ becomes a singular "surface field",
\begin{equation}
{\bf B}^b  =  B_n(u_1,u_2) H(-u_3) {\bf u}^3 - \delta(u_3) \nabla \Psi(u_1,u_2) \ .
\end{equation}
Here $\delta(x)$ is the Dirac delta function and $H(x)$ the Heaviside function.
That is, ${\bf B}^b$ is a field of finite magnetic flux in the surface, which diverts 
the normal component $B_n$ in a field along $\nabla \Psi$. This field is uniquely 
determind by the boundary condition.

\begin{figure}[h]
 \includegraphics[width=\columnwidth]{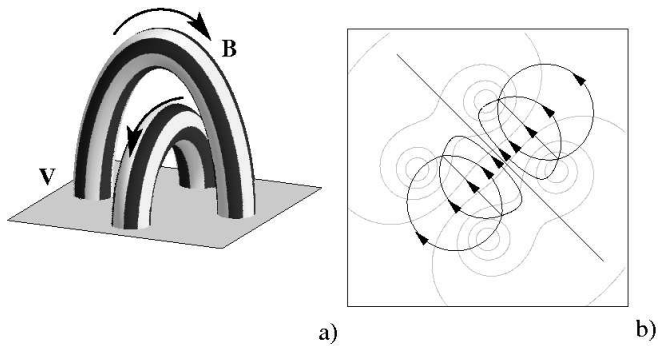}
\caption{\label{examp1} a) The example   under consideration  and b) the corresponding  boundary with 
contours of $\Psi$ (gray) and field lines of $\nabla \Psi$ (black). }
 \end{figure}

\begin{figure}[h]
 \includegraphics[width=\columnwidth]{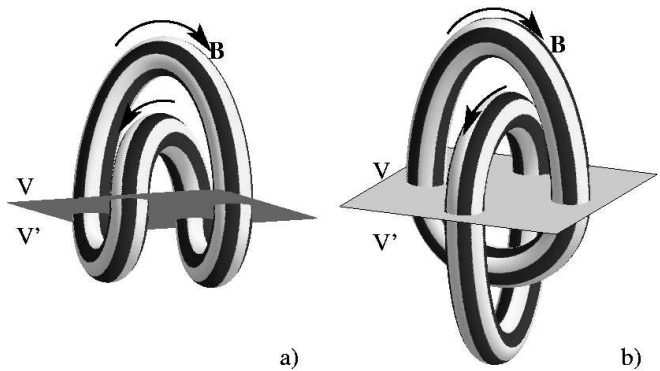}
\caption{\label{examp2} a) The field shown in Fig.~\ref{examp1} is closed by two loops which are topological equivalent to the singular boundary field. b) A closure of the field with $H_{V'}({\bf 
B})= H_{V}({\bf B})$.}
 \end{figure}

\paragraph*{Example.}For a non-trivial example, consider a field consisting of two 
untwisted flux tubes with the same magnetic flux $\Phi$ as shown in 
Fig.~\ref{examp1}(a).  There are no closed field lines in the volume under 
consideration, so there is no way of calculating the helicity of this 
configuration with the classical helicity integral. The configuration of the singular 
boundary field is shown in Fig.~\ref{examp1}(b). An explicit calculation yields  
$H_{V}({\bf B})= - \Phi^{2}$. This can be understood as the helicity of the 
closed field shown in Fig.~\ref{examp2}(a), which shows a closure of the field with a 
configuration topologically equivalent to the singular boundary field. The closed tube 
has a  uniform twist of $- 2 \pi$ and thus a total helicity of $H_{V}({\bf B})= - 
\Phi^{2}$.

Another way of calculating $H_{V}({\bf B})$ is to make use of the symmetry of the 
configuration and use Eq.~(\ref{sumvolumes}). Fig.~\ref{examp2} shows one example.
The magnetic field is closed with an identical copy in $V'$. This configuration has a 
total helicity of $- 2 \Phi^{2}$ due to the linkage of the two flux tubes 
(see e.g.~\cite{Berger1984}). 
Hence $-2 \Phi^{2} = H_{V'}({\bf B})+H_{V}({\bf B})$$= 2 H_{V}({\bf 
B})$ and therefore $H_{V}({\bf B})= - \Phi^{2}$.

\section{Summary}
In this letter it was shown how the total helicity integral can naturally be generalized to allow for magnetic fields which are not closed within the domain, i.e.~which have a non-vanishing normal component on the boundary. The construction does not require an explicit reference field as the relative helicity integral does, which was previously used in this situation. Instead we have a gauge condition for ${\bf A}$ on the boundary which corresponds to closing the domain with a topologically unique field. This field is an external complement with zero helicity density to the field in the given domain. The new integral  has all desirable properties, i.e.~it is gauge invariant, topologically invariant, and it reduces to the total helicity whenever the latter is well defined. Moreover, it shows the proper additivity with respect  to  fields  and complementary volumes. This facilitates not only many calculations of helicity, but also its interpretation.

\begin{acknowledgments} 
The author wishes to thank the Solar Theory Group in St. Andrews for their 
hospitality and acknowledges financial support  by the Royal Society of Edinburgh 
and the UK Particle Physics and Astronomy Research Council.
\end{acknowledgments}

\bibliography{hornig}
\vfill
\end{document}